\documentclass[lettersize,journal]{IEEEtran}
\usepackage{amsmath,amsfonts,amssymb,mathtools}
\usepackage{algorithmic}
\usepackage{algorithm}
\usepackage{array}
\usepackage{textcomp}
\usepackage{stfloats}
\usepackage{url}
\usepackage{graphicx}
\usepackage{booktabs}
\usepackage{multirow}
\usepackage[table]{xcolor}
\usepackage[most]{tcolorbox}
\usepackage{cite}
\usepackage[hidelinks]{hyperref}
\newtcolorbox[auto counter, number within=section]{promptbox}[2][]{%
  colback=white,
  colframe=green!50!gray!40!black,
  width=0.95\textwidth,
  arc=1mm,
  boxrule=0.5mm,
  title={\normalsize#2},
  #1
}

\begin{document}

\title{VoxAudio: Vocalized Audio Synthesis via\\ Multi-Reward Autoregressive Flow Matching}

\author{Wenxiang Guo,
        Zhou Zhao,
        Changhao Pan,
        Ziyue Jiang,
        Fei Wu%
\thanks{Manuscript received XXX; revised XXX. (Corresponding author: Fei Wu.)}%
\thanks{Wenxiang Guo, Changhao Pan, Ziyue Jiang, Zhou Zhao and Fei Wu are with Zhejiang University, Hangzhou, China (e-mail: guowx314@zju.edu.cn, panch@zju.edu.cn, ziyuejiang341@gmail.com, zhaozhou@zju.edu.cn, wufei@zju.edu.cn).}%
}

\markboth{IEEE Transactions on Multimedia}%
{Author \MakeLowercase{\textit{et al.}}: VoxAudio: Vocalized Audio Synthesis via Multi-Reward Autoregressive Flow Matching}

\maketitle

\begin{abstract}
Vocalized audio synthesis, the task of generating audio in which intelligible speech is embedded within an environmental soundscape, underpins applications such as podcast production and video dubbing. Existing Text-to-Audio (T2A) systems either reduce quoted speech to unintelligible vocal murmur or delegate it to a separate TTS model with post-hoc mixing, which forfeits control over when speech occurs and how it interacts with the scene. We present VoxAudio, a causal autoregressive flow matching model that addresses this problem from three complementary aspects. At the architecture level, chunk-wise causal factorization with independent per-chunk noise levels lets audio be emitted through sliding-window streaming inference with KV caching at variable target durations; to enable inference at arbitrary chunk granularities, we further pretrain the model with randomized chunk boundaries. At the preference level, multi-reward Negative-aware FineTuning (NFT) jointly optimizes semantic fidelity, linguistic accuracy, aesthetic quality, and temporal grounding. At the data level, to supply the missing supervision for vocal content, we build \textbf{VoxCorpus}, a large-scale corpus whose captions quote the verbatim transcript of embedded speech with time intervals, and \textbf{VoxBench}, an interval-annotated benchmark with a temporal-grounding metric. Experiments on four benchmarks spanning general audio, speech, and unified vocalized audio validate the effectiveness and efficiency of VoxAudio. Our code and demos are available at \url{https://voxaudio.github.io}.
\end{abstract}

arxiv

\section{Introduction}
\IEEEPARstart{T}{ext-to-Audio} (T2A) synthesis aims to generate high-fidelity, semantically coherent acoustic signals from natural language descriptions~\cite{huang2023make, liu2023audioldm}. As a core component of multimodal generative systems, it supports applications ranging from immersive sound design to automated content production, and has advanced rapidly from discrete autoregressive modeling~\cite{kreuk2022audiogen} to diffusion-based frameworks~\cite{yang2023diffsound, deepanway2023text}.

\begin{figure}[t]
  \centering
  \includegraphics[width=\columnwidth, trim={0cm 8cm 5cm 8cm}, clip]{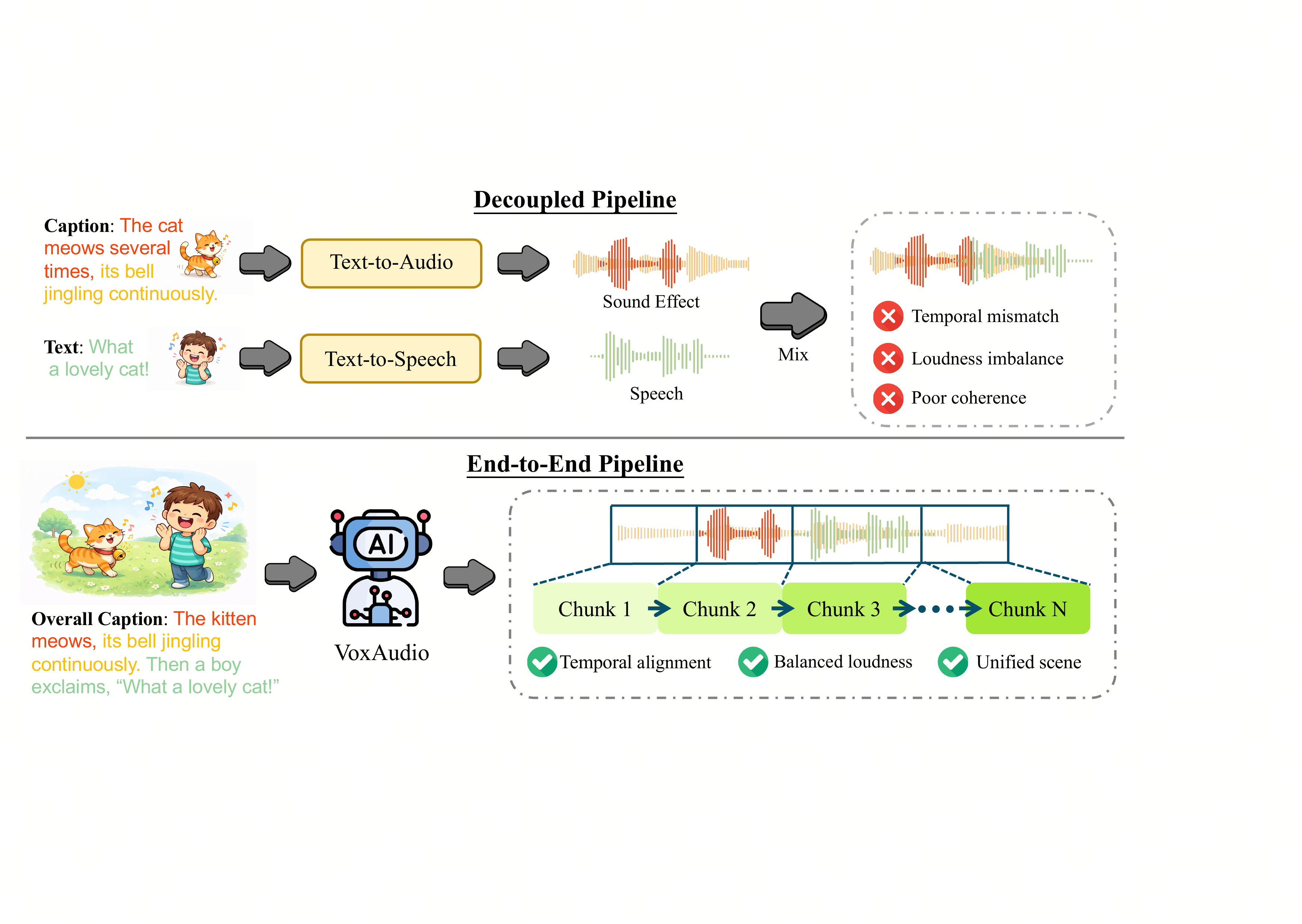}
  \caption{
    \textbf{Comparison of Audio Generation Paradigms.}
    \textit{Decoupled Pipeline}: Conventional methods synthesize soundscapes and speech independently, requiring post-hoc mixing.
    \textbf{\textit{VoxAudio}}: A unified system for vocalized audio synthesis from single prompts.
  }
  \label{fig:teaser}
\end{figure}

Extending T2A to \emph{vocalized audio}, namely audio scenes in which intelligible speech is embedded within environmental sound, nevertheless remains an open problem, and current models face three main challenges. First, most T2A models are optimized for broad acoustic events and struggle to render the lexical content specified in the caption: a quoted line is typically realized as unintelligible vocal murmur rather than the requested words. Second, although several recent works~\cite{yang2023uniaudio, vyas2023audiobox, mei2026dasheng} unify multiple audio generation tasks within a single architecture, they either treat speech and sound effects as mutually exclusive tasks or produce clips of a fixed length, offering little control over when events occur. Third, as illustrated in Figure~\ref{fig:teaser}, decoupled pipelines that synthesize speech and background sound separately and mix them post hoc cannot regulate the temporal interplay and relative loudness between the voice and ambient events, which breaks the coherence of the auditory scene.

These challenges can be traced to causes at three levels. At the data level, existing corpora rarely annotate environmental sounds and verbatim speech jointly within a single description, leaving models without direct supervision for text-to-speech mappings inside complex acoustic contexts. At the architecture level, mainstream audio generators adopt non-autoregressive formulations that synthesize a fixed-length clip in a single pass, which supports neither streaming output nor a natural interface for variable target durations. At the level of the training paradigm, purely supervised likelihood training draws generation toward the average of the data and offers no direct handle on human preferences such as intelligibility, semantic fidelity, and perceptual quality.

To address these challenges, we present \textbf{VoxAudio} together with the data it learns from. At the data level, we construct \textbf{VoxCorpus}, a large-scale corpus that pairs a controlled simulation pipeline, which composes speech and environmental sounds under exact timing and loudness, with real narrative recordings annotated by a transcribe-then-fuse captioning pipeline; every caption quotes the verbatim transcript of the embedded speech together with its time interval. At the architecture level, VoxAudio is a causal autoregressive flow matching model: the latent sequence is partitioned into causal chunks that carry independent noise levels during training, and the backbone is pretrained with \emph{randomized chunk boundaries} rather than a fixed chunk size, so the streaming granularity can be chosen freely at inference. At inference, a sliding window advances adjacent chunks with a fixed denoising lag under key-value caching, emitting audio continuously while later chunks are still being refined. At the level of the training paradigm, a multi-reward Negative-aware FineTuning (NFT)~\cite{zheng2025diffusionnft} stage adapted to causal flow matching jointly optimizes semantic fidelity, linguistic accuracy measured by word error rate, perceptual aesthetics, and temporal grounding.
To evaluate the generated vocalized audio, we further build \textbf{VoxBench}, an interval-annotated benchmark equipped with a temporal-grounding metric, and compare VoxAudio with strong baselines on general text-to-audio, text-to-speech, and unified vocalized audio generation under a unified protocol.

Our primary contributions are summarized as follows:
\begin{itemize}
    \item We propose a streamable autoregressive flow matching architecture with chunk-agnostic causal factorization; by pretraining over randomized chunk boundaries the model supports an arbitrary streaming chunk size chosen during inference, providing low-latency streaming and variable-duration generation.
    \item We introduce a multi-reward NFT post-training stage for preference alignment that jointly optimizes semantic fidelity, linguistic accuracy, aesthetic quality, and temporal grounding.
    \item We construct VoxCorpus, a large-scale corpus with quoted, time-stamped speech annotations, and VoxBench, an interval-annotated evaluation benchmark equipped with a temporal-grounding metric.
    \item Empirical results show that VoxAudio outperforms current vocalized audio generation baselines in speech fidelity and timing precision and stays competitive with dedicated T2A and TTS systems.
\end{itemize}

\section{Related Works}

\subsection{Text-to-Audio Synthesis}
Text-to-Audio (T2A) synthesis aims to generate controllable, high-fidelity audio conditioned on textual descriptions.
Early works cast the problem as discrete sequence modeling: DiffSound~\cite{yang2023diffsound} generates mel-spectrogram tokens with a discrete diffusion decoder, while AudioGen~\cite{kreuk2022audiogen} autoregressively predicts neural-codec tokens from descriptive captions.
Latent Diffusion Models (LDMs) subsequently became the dominant paradigm owing to their generation quality: Make-An-Audio~\cite{huang2023make}, AudioLDM~\cite{liu2023audioldm}, and TANGO~\cite{deepanway2023text} denoise in the latent space of a variational autoencoder (VAE) under CLAP- or LLM-based text conditioning, and follow-ups strengthen temporal ordering through structured captions \cite{huang2023make2}, scale data and backbones~\cite{liu2024audioldm, haji2024taming}, and adopt diffusion transformers with flow-matching objectives for faster, higher-fidelity synthesis~\cite{hai2024ezaudio, evans2025stable, cheng2025mmaudio}.
Beyond single-task synthesis, unified frameworks such as UniAudio~\cite{yang2023uniaudio}, Audiobox~\cite{vyas2023audiobox}, and AudioX~\cite{tian2025audiox} integrate multiple audio generation tasks into one model, and the concurrent Dasheng-AudioGen~\cite{mei2026dasheng} scales a non-causal DiT with structured multi-view captions toward unified speech, music, and sound generation.
Despite this progress, two limitations persist.
First, these methods struggle to render the exact linguistic content quoted in captions: speech is either excluded by design or synthesized as a generic vocal texture, since none of them receives verbatim transcript supervision inside scene-level descriptions.
Second, diffusion-based systems operate on global sequences of fixed length, requiring the entire clip to be generated simultaneously; this non-causal formulation precludes streaming and ties the model to a fixed clip length, limiting real-time interactive use.
VoxAudio couples a causal autoregressive formulation, which makes streaming and variable-duration generation possible.

\subsection{Text-to-Speech Synthesis}
Modern text-to-speech (TTS) systems achieve near-human intelligibility by scaling language-model or flow-matching architectures on large speech corpora. Codec language models such as \cite{wang2023neural} treat TTS as discrete token prediction conditioned on a reference speaker prompt, while non-autoregressive approaches such as \cite{le2023voicebox} and \cite{ju2024naturalspeech} employ flow matching or factorized codecs for faster inference. F5-TTS~\cite{chen2025f5} shows that a text-conditioned flow-matching transformer alone suffices without explicit duration modeling; CosyVoice~2/3~\cite{du2024cosyvoice, du2025cosyvoice} and Seed-TTS~\cite{anastassiou2024seed} combine LLM front-ends with streaming synthesizers for scalable in-the-wild generation, and recent diffusion-hybrid designs~\cite{jia2025ditar, peng2025vibevoice} push long-form expressive synthesis.
However, these systems synthesize clean, foreground speech in isolation: environmental context must be added by a separate T2A model and a mixing stage, which severs the acoustic coupling between voice and scene. Our work instead targets the joint distribution of speech and environment within a single generative process.

\subsection{Autoregressive Diffusion}
Combining the sequential priors of autoregressive models with the continuous modeling power of diffusion has attracted growing interest.
In the image domain, \cite{li2024autoregressive} replaces discrete tokens with per-token diffusion heads, and hybrid decoders~\cite{sun2024multimodal} interleave autoregressive planning with diffusion refinement.
In the video domain, Diffusion Forcing~\cite{chen2024diffusion} trains with per-frame independent noise levels to enable causal rollout, AR-Diffusion~\cite{sun2025ar} designs asynchronous timestep schedulers, and streaming generators~\cite{yin2025slow} distill bidirectional teachers into causal students for real-time synthesis.
In the speech domain, DiTAR~\cite{jia2025ditar} and related studies~\cite{liu2024autoregressive, peng2025vibevoice} generate continuous latents chunk-by-chunk with diffusion decoding heads.
Our work adapts the asynchronous-noise principle to general audio with two distinctions: chunk boundaries are \emph{randomized} during pretraining so that the learned dynamics are agnostic to the streaming granularity, and the training-time noise-level patterns are matched to the sliding-window schedule used at inference, closing the train-test gap that fixed-chunk causal models exhibit.

\subsection{Preference Alignment for Generative Models}
Reinforcement learning from human feedback~\cite{ouyang2022training} and its direct variants~\cite{rafailov2023direct, shao2024deepseekmath} have become standard for aligning large language models, and analogous techniques now align visual generators: Diffusion-DPO~\cite{wallace2024diffusion} transfers preference optimization to denoising trajectories, while Flow-GRPO~\cite{liu2025flow} 
applys group-relative policy optimization to flow-matching models.
In the audio domain, Tango2~\cite{majumder2024tango} constructs preference pairs for DPO, TangoFlux~\cite{hung2024tangoflux} introduces CLAP-ranked preference optimization, and PrismAudio~\cite{liu2025prismaudio} accelerates GRPO with hybrid ODE-SDE sampling; most of these efforts optimize a single semantic reward.
Most recently, DiffusionNFT~\cite{zheng2025diffusionnft} optimizes diffusion models directly on the forward process via a negative-aware flow-matching objective, avoiding likelihood estimation and solver backpropagation.
We extend this paradigm along two axes: the objective is adapted to causal flow matching by replaying the asynchronous noise snapshots of the streaming sampler, and the scalar reward is replaced by a weighted combination of semantic, linguistic, aesthetic, and temporal signals.

\section{VoxAudio}
\label{sec:method}
\begin{figure*}[t]
  \begin{center}
    \centerline{\includegraphics[width=\textwidth, trim={1.5cm 1cm 3.5cm 2.5cm}, clip]{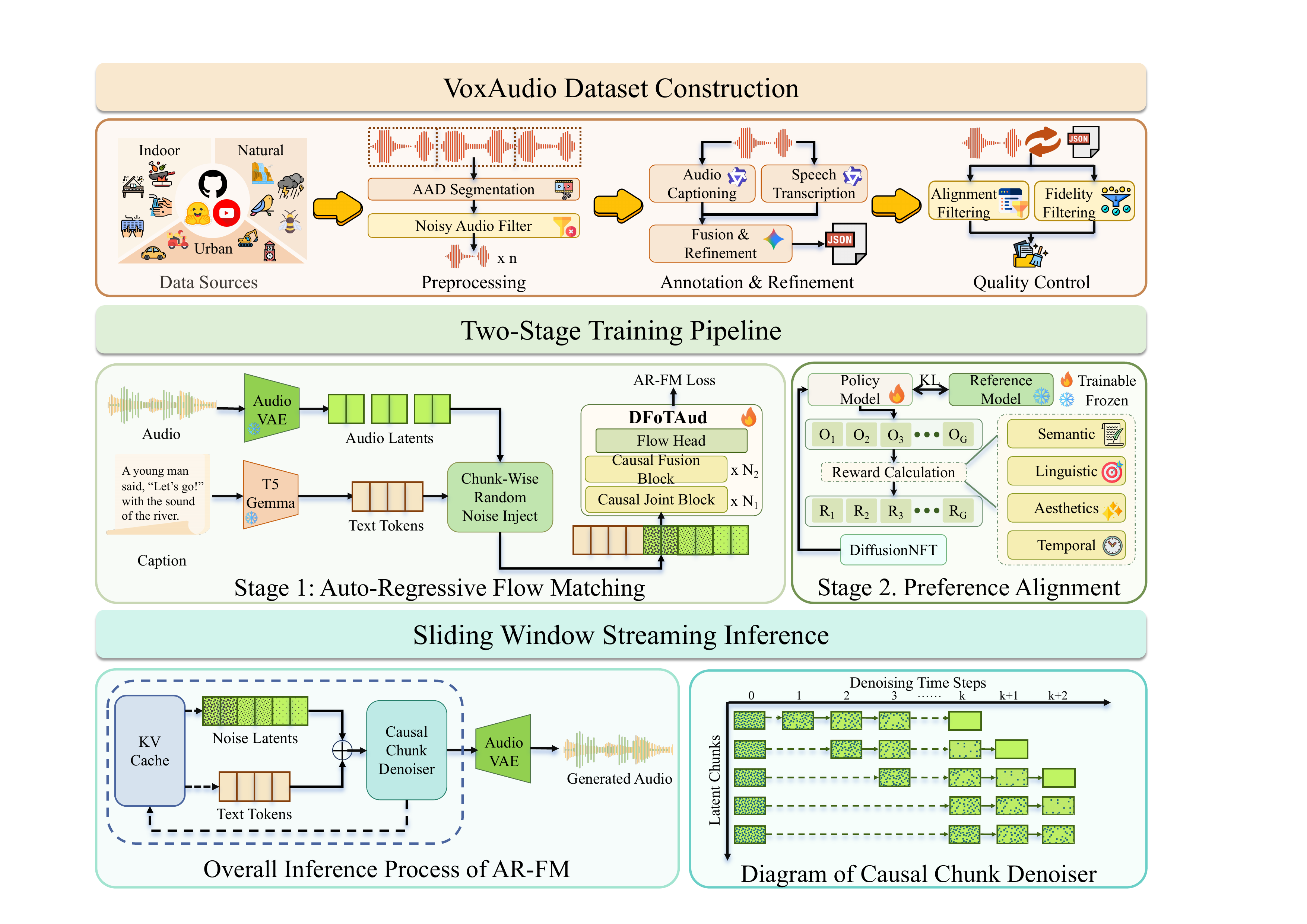}}
    \caption{
      \textbf{Overview of the VoxAudio Framework.}
      \textbf{Top:} The workflow for VoxAudio dataset construction.
      \textbf{Middle:} The training pipeline, comprising chunk-agnostic autoregressive flow-matching supervised training and preference alignment NFT with multi-dimensional rewards.
      \textbf{Bottom:} The sliding-window streaming inference mechanism, which facilitates real-time, continuous audio synthesis.
    }
    \label{fig:overall_arch}
  \end{center}
  \vspace{-0.6em}
\end{figure*}

This section presents the VoxAudio framework. Section~\ref{sec:ar_fm} formulates autoregressive flow matching (AR-FM), which assigns independent noise levels to causal chunks. Section~\ref{sec:model_arch} describes the model architecture, covering the Universe audio VAE, textual condition injection, and the causal adaptations of the backbone, namely causal convolutions, chunk-wise causal attention, and duration conditioning. Section~\ref{sec:rl} presents the multi-reward preference alignment objective. Section~\ref{sec:training} summarizes the two-stage training pipeline that combines these components, and Section~\ref{sec:inference} details the sliding-window streaming inference procedure.

\subsection{Autoregressive Flow Matching}
\label{sec:ar_fm}
VoxAudio aims to model the conditional distribution $p(x|c)$ of a high-fidelity audio waveform $x \in \mathbb{R}^{T}$ given textual and duration conditions $c$, where $T$ denotes the temporal duration. Utilizing a pre-trained VAE, the audio is encoded into a latent representation $\mathbf{z} \in \mathbb{R}^{L \times D}$, where $L$ represents the number of frames and $D$ is the channel dimension.

Standard flow matching defines a probability density path $p_t(\mathbf{z})$ that continuously deforms a Gaussian prior $p_0(\mathbf{z}) = \mathcal{N}(\mathbf{z}; 0, I)$ to the data distribution $p_1(\mathbf{z}) \approx p_{\text{data}}(\mathbf{z})$ under a synchronous global timestep $t$. This assumption contradicts streaming generation, where historical frames are fully denoised while future frames remain purely noisy. To bridge this gap, as illustrated in Figure~\ref{fig:overall_arch}, our AR-FM strategy adapts asynchronous denoising dynamics to the audio latent space: we partition the latent sequence $\mathbf{z}$ along the temporal dimension into $K$ chunks $\mathcal{Z} = \{\mathbf{z}^{(1)}, \mathbf{z}^{(2)}, \dots, \mathbf{z}^{(K)}\}$. Instead of a global timestep, we sample a timestep sequence $\boldsymbol{t} = \{t_1, t_2, \dots, t_K\}$, where $t_i \in [0, 1]$ represents the noise level for the $i$-th chunk.
We adopt the conditional flow matching paradigm with an optimal transport path to ensure straight trajectories, thereby improving sampling efficiency. For a data sample $\mathbf{z}^{(i)}_1$ and noise $\mathbf{z}^{(i)}_0$, the intermediate state is defined linearly:
\begin{equation}
    \mathbf{z}^{(i)}_{t_i} = (1 - t_i) \mathbf{z}^{(i)}_0 + t_i \mathbf{z}^{(i)}_1,
\end{equation}
yielding the target vector field $u_t(\mathbf{z}^{(i)}|\mathbf{z}^{(i)}_1) = \mathbf{z}^{(i)}_1 - \mathbf{z}^{(i)}_0$. The training objective is to minimize the Mean Squared Error between the predicted vector field $v_\theta$ and the target:
\begin{equation}
    \mathcal{L}_{\text{FM}} = \sum_{i=1}^{K} \mathbb{E}_{\boldsymbol{t}, \mathbf{z}_0, \mathbf{z}_1} \left[ \left\| u_{t_i} - v_\theta\big(\mathbf{z}^{(i)}_{t_i} \,\big|\, t_i, c, \{\mathbf{z}^{(j)}_{t_j}\}_{j<i}\big) \right\|^2 \right],
\end{equation}
where $\{\mathbf{z}^{(j)}_{t_j}\}_{j<i}$ denotes all preceding chunks at their own noise levels: the prediction for chunk $i$ attends to them through the chunk-wise causal attention, so the factorization is autoregressive across chunks rather than independent per chunk. Because the per-chunk noise levels are sampled independently during training, the model observes every configuration of partially denoised history and noisy future, so the asynchronous states visited by streaming inference lie within the training distribution rather than constituting a train-test gap.

\begin{figure*}[t]
  \centering
  \includegraphics[width=0.92\textwidth, trim={0cm 5cm 0cm 1cm}, clip]{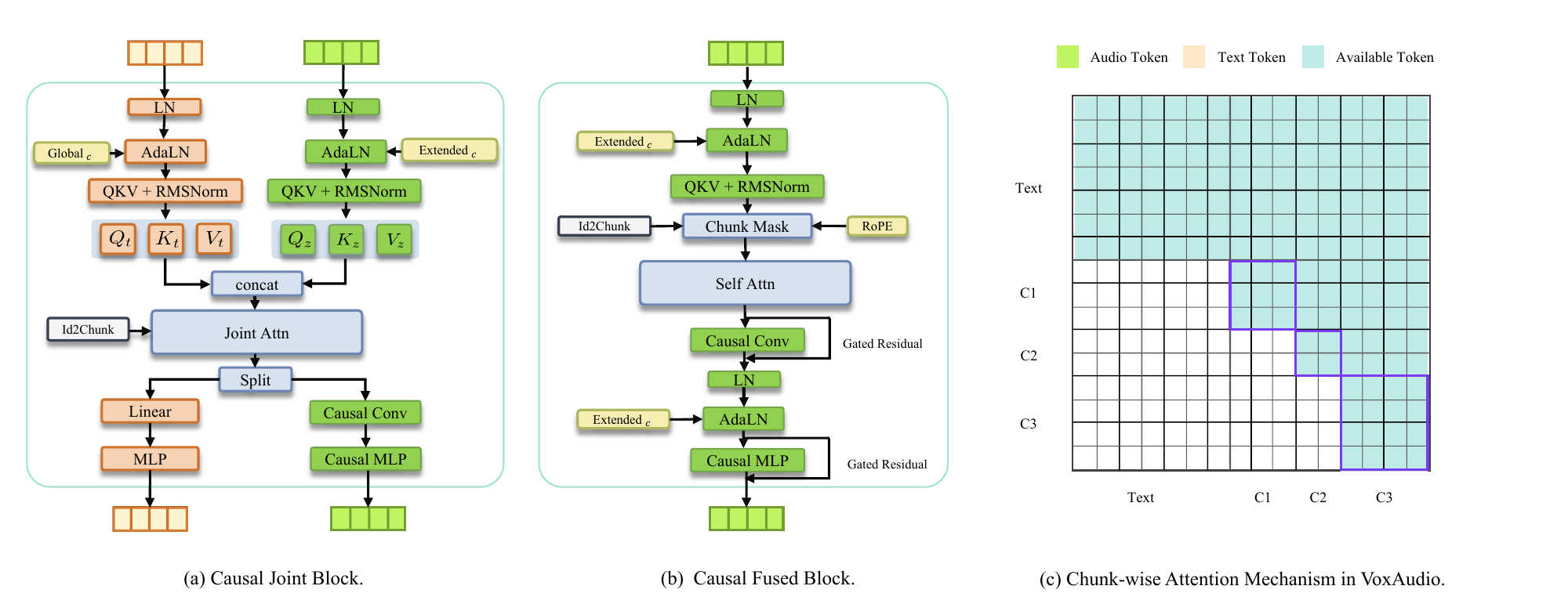}
  \caption{
    \textbf{Detailed architecture and attention mechanism of the Causal Diffusion Transformer.}
    (a) The causal joint block facilitates interaction between audio latents and conditioning features;
    (b) The causal fused block refines integrated representations through deep self-attention;
    (c) The chunk-wise causal attention mask, where shaded regions indicate masked-out tokens, ensuring that each temporal chunk only attends to current and historical context.
  }
  \label{fig:trans}
\end{figure*}

\subsection{Model Architecture}
\label{sec:model_arch}

As shown in Figure~\ref{fig:overall_arch}, we build our audio foundation model upon the diffusion transformer backbone~\cite{cheng2025mmaudio}. To adapt this backbone for causal inference, we introduce specific architectural modifications regarding condition injection and attention mechanisms.

\textbf{Universe Audio VAE.}
VoxAudio operates in the latent space of the pre-trained Universe audio VAE, which compresses the waveform into a low-frame-rate latent sequence. The VAE is trained on a broad mixture of speech, sound effects, and music, so a single latent space faithfully represents both articulate vocal content and diverse acoustic scenes, which is a prerequisite for unified vocalized audio generation.

\textbf{Textual Condition Injection.}
To robustly extract semantic features, we employ a frozen T5-family text encoder~\cite{zhang2025encoder} and concatenate features from multiple encoder depths (first, middle, and last layers), capturing both the lexical surface of the quoted speech and the abstract semantics of the scene description, including interval annotations.

\textbf{Causal Convolutions.}
Standard convolutions pad symmetrically and thus leak future context, which violates the causality required for streaming. We replace all convolutions in the backbone with their causal, left-padded counterparts, so that the output at frame $l$ depends solely on frames $l' \le l$; at streaming inference, each layer carries a small state buffer across chunks, making chunk-by-chunk computation identical to full-sequence computation and free of boundary artifacts.

\textbf{Chunk-wise Causal Attention.}
Following the MMDiT architecture~\cite{esser2024scaling}, we represent the input as a concatenated sequence $\mathbf{z} = [\mathbf{z}_{\text{txt}}, \mathbf{z}_{\text{lat}}]$ of length $N = L_{\text{txt}} + L_{\text{lat}}$. To balance global causality with local coherence, we implement a Chunk-wise Causal Attention mechanism. Unlike standard causal masks, this approach permits bidirectional visibility within each chunk while strictly enforcing causality across chunk boundaries. Let $g(k)$ be the chunk index function mapping each global token index $k$ to its block ($g(k)=0$ for text tokens). The attention mask $M_{i,j}$ for a query at index $i$ and a key at index $j$ is defined such that a token can attend to all tokens within its own chunk and all preceding chunks:
\begin{equation}
M_{i,j} =
\begin{cases}
0, & \text{if } g(j) \le g(i) \\
-\infty, & \text{otherwise}
\end{cases}.
\end{equation}
This configuration ensures that the text prompt is fully visible to all subsequent audio latents and that each audio frame can leverage the full context of its current chunk. Figure~\ref{fig:trans} details the two block types and the resulting attention mask.

\textbf{Duration Conditioning.}
To provide an explicit stop signal for streaming generation and precise control over the output length, the target duration is encoded by a Fourier feature encoder and added to the global condition; the duration condition is randomly dropped during training to enhance robustness.

\subsection{Multi-Reward Preference Alignment}
\label{sec:rl}
To align the flow-matching objective with human-centric preferences across acoustic, semantic, and aesthetic dimensions, we introduce a multi-reward reinforcement learning framework. Inspired by DiffusionNFT~\cite{zheng2025diffusionnft}, we adapt this paradigm to the AR-FM framework. DiffusionNFT enables optimization using arbitrary reward functions without requiring gradient backpropagation through the solver, making it ideally suited for our AR-FM architecture.

\subsubsection{Weighted Multi-dimensional Reward}
Synthesizing high-fidelity audio requires a holistic evaluation beyond simple TTA reconstruction. We define a composite reward comprising four dimensions. \emph{Semantic fidelity}: LAION-CLAP~\cite{laionclap2023} audio-text similarity measures the alignment between the rollout and the caption; to avoid reward hacking on this dimension, we adopt PE\textsubscript{AV}~\cite{vyas2025pushingfrontieraudiovisualperception} as the semantic metric in our experiments. \emph{Linguistic accuracy}: Whisper~\cite{radford2023robust} computes the Word Error Rate (WER). \emph{Perceptual aesthetics}: the Meta Audiobox Aesthetics model~\cite{tjandra2025aes} guides the generation toward high-quality soundscapes through its production-quality, content-enjoyment, and content-usefulness axes. \emph{Temporal grounding}: the intersection-over-union between the speech spans detected by PE-A-Frame~\cite{bolya2025PerceptionEncoder} and the interval annotated in the caption. Together, the four rewards constrain what is said, when it is said, and how the scene sounds.

The four dimension scores are aggregated into a single scalar training signal through group-wise rollouts. For each caption $c$, we generate a group of $K$ samples $\{\hat{\mathbf{x}}_k\}$ with the frozen rollout policy; the individual scores are balanced by pre-defined weights $\{w_i\}$ into a scalar reward, the per-sample advantage is normalized within the group, and clipping the scaled advantage yields the final reward $r \in [0, 1]$. To prevent the policy from collapsing to a single clip length, RL prompts specify \emph{variable durations}, bucketed per batch, so preference optimization is exercised across the whole duration range.

\subsubsection{Implicit Policy Optimization}
Building upon the Negative-aware FineTuning (NFT) paradigm, we adapt the optimization process to the flow-matching latent space. We define the update direction as the velocity deviation $\Delta v = v_\theta - v_{\text{old}}$, representing the refinement of the vector field. To steer the policy towards high-reward regions while avoiding suboptimal trajectories, we construct implicit positive and negative target policies formulated as $v_{\text{old}} + \beta \Delta v$ and $v_{\text{old}} - \beta \Delta v$, respectively. We formulate the Negative-aware FineTuning loss as:
\begin{equation}
\begin{split}
\mathcal{L}_{\text{NFT}} = \frac{1}{\beta} \mathbb{E} \big[ & r \|(v_{old} + \beta \Delta v) - v_{target}\|^2 + \\
& (1-r) \|(v_{old} - \beta \Delta v) - v_{target}\|^2 \big],
\end{split}
\end{equation}
where $\beta$ is a scaling factor. Two adaptations are specific to our causal setting. First, training timesteps are not sampled randomly per chunk. Instead, we replay the exact asynchronous noise-level snapshots that the streaming sampler visits: for a uniformly sampled global solver step $g$, the noise level assigned to chunk $k$ is
\begin{equation}
t_k(g) = \operatorname{clip}\big( (g - (k-1)\Delta)\,\delta,\ 0,\ 1 \big),
\end{equation}
where $\Delta$ is the streaming step lag and $\delta$ the solver step size, matching Algorithm~\ref{alg:infer_ar} exactly. The fine-tuned vector field is therefore optimized precisely on the states encountered at inference. Second, the rollout anchor $v_{\text{old}}$ is softly updated toward the policy with a ramped rate, stabilizing on-policy improvement. To preserve the model's generalization capability and prevent catastrophic forgetting of the pre-trained distribution, we include a KL-regularization term $\mathcal{L}_{\text{KL}} = \|v_\theta - v_{ref}\|^2$ against a frozen reference. 

\subsection{Two-Stage Training Pipeline}
\label{sec:training}
As shown in Figure~\ref{fig:overall_arch}, VoxAudio is trained in two stages, supervised flow-matching training followed by multi-reward preference alignment.

\textbf{Supervised Flow-Matching Training.}
The model is first trained from scratch on the timed simulation corpus with the AR-FM objective. The chunk partition is central to streamability: a single fixed chunk size would entangle the learned representation with one particular streaming granularity, whereas the degenerate case of a single chunk covering the whole clip reduces the model to a non-autoregressive generator. Instead of committing to either extreme, we sample the chunk partition itself: every latent frame independently opens a new chunk with probability $p_b$, producing chunks whose lengths follow a geometric distribution with mean $1/p_b$ frames. The model therefore observes a continuum of granularities and learns denoising dynamics that are agnostic to the chunk layout, so the streaming chunk size can be chosen freely during inference. Section~\ref{sec:ablation} shows that this chunk-agnostic model, run at a fixed inference chunk size, stays on par with backbones fine-tuned for that specific size, while the single-chunk variant degrades linguistic accuracy and cannot stream. Within the same supervised stage, training then continues on a blend of the real narrative corpus and the simulation corpus, which adapts the model to in-the-wild acoustic conditions and narrative caption styles.

\textbf{Multi-Reward Preference Alignment.}
The model then undergoes the NFT stage with variable-duration rollouts, aligning semantic fidelity, linguistic accuracy, perceptual aesthetics, and temporal grounding with listener preferences.

\begin{algorithm}[t]
   \caption{Sliding Window Inference for AR-FM}
   \label{alg:infer_ar}
\begin{algorithmic}[1]
   \STATE {\bfseries Input:} Condition $c$, $K$ chunks, steps $N$, lag $\Delta$, step size $\delta = 1/N$.
   \STATE {\bfseries Initialize:} $\mathbf{z}^{(1:K)} \sim \mathcal{N}(0, \mathbf{I})$, $\text{KV-Cache } \mathcal{K} = \emptyset$.
   \FOR{$g = 0$ {\bfseries to} $N + (K-1)\Delta - 1$}
       \STATE $\mathcal{I}_{\text{act}} = \{k \mid 1 \le k \le K \text{ and } 0 \le t_k < 1,\ t_k = (g - (k-1)\Delta)\delta\}$
       \FOR{{\bfseries each} $k \in \mathcal{I}_{\text{act}}$}
           \STATE $\mathbf{z}^{(k)} \leftarrow \mathbf{z}^{(k)} + \delta \cdot v_\theta(\mathbf{z}^{(k)}, t_k, c, \mathcal{K})$
           \IF{$t_k + \delta \approx 1$}
               \STATE Stream $\text{Decode}(\mathbf{z}^{(k)})$ and update $\mathcal{K}$.
           \ENDIF
       \ENDFOR
   \ENDFOR
\end{algorithmic}
\end{algorithm}

\subsection{Sliding Window Streaming Inference}
\label{sec:inference}
A straightforward streaming strategy denoises chunks strictly sequentially: the first chunk $\mathbf{z}^{(1)}$ is fully denoised, its key-value states are cached, and generation then proceeds to $\mathbf{z}^{(2)}$. Although this schedule minimizes the latency of the first packet, it leaves all subsequent chunks idle during each denoising pass and therefore yields low throughput and an extended total generation time.

To address this, as illustrated in Figure~\ref{fig:overall_arch}, we propose a Sliding Window Inference strategy, detailed in Algorithm~\ref{alg:infer_ar}. Instead of generating one chunk at a time, we process a window of active chunks simultaneously. These chunks maintain a fixed timestep difference: when chunk $i$ is at denoising step $s$, chunk $i+1$ is at step $s-\Delta$. This approach balances initial latency with overall throughput, and a moderate step difference keeps the noise-level pattern close to the asynchronous schedules observed during training. Classifier-free guidance is applied on the text condition, using two batched streams that share the key-value cache of the already-generated history:
\begin{equation}
\hat{v} = v_\theta(\mathbf{z}, t, \varnothing, \mathcal{K}) + s \cdot \big( v_\theta(\mathbf{z}, t, c, \mathcal{K}) - v_\theta(\mathbf{z}, t, \varnothing, \mathcal{K}) \big),
\end{equation}
where $s$ is the guidance scale. Sampling uses the exponential-moving-average (EMA) weights of the policy.

\section{Dataset and Benchmark}
\label{sec:dataset}
\begin{table*}[t]
\caption{Unified vocalized audio generation on VoxBench-10s and MECAT-en.}
\label{tab:unified}
\begin{center}
\begin{small}
\setlength{\tabcolsep}{4pt}
\begin{tabular}{lccccccccc}
\toprule
& \multicolumn{2}{c}{Semantic} & Linguistic & \multicolumn{3}{c}{Aesthetics} & Temporal & \multicolumn{2}{c}{Subjective} \\
\cmidrule(lr){2-3} \cmidrule(lr){4-4} \cmidrule(lr){5-7} \cmidrule(lr){8-8} \cmidrule(lr){9-10}
Model & PE\textsubscript{AV}$\uparrow$ & CLAP$\uparrow$ & WER$\downarrow$ & PQ$\uparrow$ & CE$\uparrow$ & CU$\uparrow$ & TG-IoU$\uparrow$ & MOS-O$\uparrow$ & MOS-C$\uparrow$ \\
\midrule
\multicolumn{10}{l}{\textit{(a) VoxBench-10s}} \\
\midrule
Dasheng-AudioGen~\cite{mei2026dasheng} & 0.110 & 0.396 & 0.264 & 5.94 & 4.10 & 5.15 & 0.146 & 4.42{\scriptsize$\pm$0.18} & 4.16{\scriptsize$\pm$0.34} \\
w/o Semantic Reward   & 0.085 & 0.250 & 0.042 & 7.37 & 4.45 & 6.61 & 0.825 & 4.40{\scriptsize$\pm$0.17} & 4.34{\scriptsize$\pm$0.28} \\
w/o Aesthetic Reward  & 0.119 & 0.391 & 0.054 & 5.77 & 4.08 & 5.21 & 0.633 & 4.45{\scriptsize$\pm$0.12} & 4.49{\scriptsize$\pm$0.23} \\
w/o Linguistic Reward & 0.118 & 0.398 & 0.056 & 6.05 & 4.21 & 5.39 & 0.652 & 4.48{\scriptsize$\pm$0.13} & 4.67{\scriptsize$\pm$0.18} \\
w/o Temporal Reward   & 0.118 & 0.391 & 0.044 & 5.85 & 4.13 & 5.29 & 0.609 & 4.37{\scriptsize$\pm$0.17} & 4.59{\scriptsize$\pm$0.22} \\
w/o All Rewards       & 0.113 & 0.353 & 0.081 & 5.61 & 4.07 & 5.15 & 0.652 & 4.43{\scriptsize$\pm$0.14} & 4.66{\scriptsize$\pm$0.20} \\
\textbf{VoxAudio (ours)} & 0.116 & 0.392 & 0.038 & 5.96 & 4.21 & 5.39 & 0.654 & 4.47{\scriptsize$\pm$0.13} & 4.51{\scriptsize$\pm$0.26} \\
\midrule
\multicolumn{10}{l}{\textit{(b) MECAT-en}} \\
\midrule
Dasheng-AudioGen~\cite{mei2026dasheng} & 0.125 & 0.466 & 0.151 & 6.60 & 5.18 & 6.10 & - & 4.25{\scriptsize$\pm$0.13} & 4.57{\scriptsize$\pm$0.21} \\
w/o Semantic Reward   & 0.092 & 0.324 & 0.081 & 7.00 & 5.01 & 6.70 & - & 4.24{\scriptsize$\pm$0.23} & 4.45{\scriptsize$\pm$0.26} \\
w/o Aesthetic Reward  & 0.116 & 0.436 & 0.086 & 6.00 & 4.56 & 5.62 & - & 4.24{\scriptsize$\pm$0.17} & 4.50{\scriptsize$\pm$0.26} \\
w/o Linguistic Reward & 0.118 & 0.447 & 0.087 & 6.25 & 4.75 & 5.84 & - & 4.22{\scriptsize$\pm$0.17} & 4.31{\scriptsize$\pm$0.29} \\
w/o Temporal Reward   & 0.118 & 0.454 & 0.089 & 6.51 & 4.87 & 6.11 & - & 4.33{\scriptsize$\pm$0.16} & 4.46{\scriptsize$\pm$0.23} \\
w/o All Rewards       & 0.125 & 0.449 & 0.088 & 6.05 & 4.68 & 5.74 & - & 4.10{\scriptsize$\pm$0.21} & 4.30{\scriptsize$\pm$0.29} \\
\textbf{VoxAudio (ours)} & 0.123 & 0.450 & 0.085 & 6.30 & 4.76 & 5.94 & - & 4.23{\scriptsize$\pm$0.22} & 4.55{\scriptsize$\pm$0.23} \\
\bottomrule
\end{tabular}
\end{small}
\end{center}
\end{table*}

To address the scarcity of high-fidelity audio datasets that integrate both complex environmental soundscapes and explicit speech content, we curate \textbf{VoxCorpus}, which combines a controlled simulation pipeline with a real narrative-audio collection, and develop \textbf{VoxBench}, a comprehensive benchmark for reinforcement learning and evaluation. Detailed data curation is provided in the supplementary material.

\subsection{VoxCorpus: Training Dataset}
Existing public datasets, such as AudioCaps \cite{kim2019audiocaps} and WavCaps~\cite{mei2024wavcaps}, primarily focus on general acoustic categories and provide neither explicit transcriptions for the embedded speech content nor annotations of its temporal extent. We therefore build two complementary corpora, accompanied by a dedicated prompt set for preference alignment.

\textbf{Timed Simulation Corpus.}
We programmatically compose clean speech utterances with non-speech recordings under five layout modes: speech-only, sound-only, speech-then-sound, sound-then-speech, and overlay. The mixer controls the speech onset within the clip, the speech-to-background signal-to-noise ratio, and an optional trailing-silence segment. Each composition is rendered together with a structured caption that quotes the transcript verbatim and annotates segment intervals, e.g., ``\textit{Total duration 10s. Someone says: `...' (from 3.5s to 5.5s), while simultaneously the sound of heavy rain on a tin roof (from 0s to 10s).}'' Total-duration prefixes, segment times, and numeric precision are stochastically dropped so the model accepts captions at any annotation density. 

\textbf{Real Narrative Corpus.}
As illustrated in the top panel of Figure~\ref{fig:overall_arch}, we curate a speech-integrated real corpus through three stages.
\emph{Data aggregation and preprocessing}: we aggregate raw audio from diverse open-source repositories and extensive web crawling, and apply an energy-based Audio Activity Detection pipeline\footnote{\url{https://github.com/amsehili/auditok}} to segment long recordings into clips under 15 seconds.
\emph{Hierarchical annotation and refinement}: following a ``transcribe-then-synthesize'' strategy, Qwen3-Omni~\cite{Qwen3-Omni} generates environmental captions, Qwen3-ASR transcribes embedded speech, and Gemini~\cite{comanici2025gemini} fuses both into narrative captions that explicitly quote vocalizations in temporal order.
\emph{Data stratification}: we filter by LAION-CLAP~\cite{laionclap2023} alignment and the PE\textsubscript{AV}~\cite{vyas2025pushingfrontieraudiovisualperception} perceptual score, yielding VoxCorpus-Curated, a distilled subset of 879{,}768 high-fidelity samples.

\textbf{RL Prompt Set.}
Preference alignment uses a dedicated set of 4,261 prompts with target durations of 10--30\,s: 1,876 mixed speech-plus-scene prompts, 936 speech-only prompts, and 1,449 sound-only prompts. They are sampled from the original captions of the AudioCaps training split and real-corpus captions.

\subsection{VoxBench: Evaluation Benchmark}
To evaluate instruction following for vocalized audio and to provide prompts for reinforcement learning, we construct VoxBench. It consists of LLM-synthesized audio scenes spanning 12 acoustic domains and a curated library of fine-grained keywords. For each benchmark item, we randomly sample one scene and one keyword, and generate a caption that describes the scenario while explicitly incorporating the requested vocal content. We then manually review all synthesized instances to remove implausible or low-quality samples, resulting in 977 held-out items.

\begin{table*}[t]
\caption{Quantitative results on the AudioCaps test set. }
\label{tab:audiocaps}
\begin{center}
\begin{small}
\setlength{\tabcolsep}{4pt}
\resizebox{\textwidth}{!}{%
\begin{tabular}{lccccccccccccc}
\toprule
& & \multicolumn{4}{c}{Acoustic} & \multicolumn{2}{c}{Semantic} & \multicolumn{3}{c}{Aesthetics} & \multicolumn{2}{c}{Subjective} & Efficiency \\
\cmidrule(lr){3-6} \cmidrule(lr){7-8} \cmidrule(lr){9-11} \cmidrule(lr){12-13} \cmidrule(lr){14-14}
Model & Params(M) & FD-VGG$\downarrow$ & FD-PANN$\downarrow$ & ISC$\uparrow$ & KL$\downarrow$ & CLAP$\uparrow$ & PE\textsubscript{AV}$\uparrow$ & PQ$\uparrow$ & CE$\uparrow$ & CU$\uparrow$ & MOS-O$\uparrow$ & MOS-C$\uparrow$ & RTF$\downarrow$ \\
\midrule
GenAU~\cite{haji2024taming}                & 1250 & 1.50 & 29.5 & 10.7 & 1.42 & 0.635 & 0.135 & 5.70 & 3.39 & 4.95 & 4.21{\scriptsize$\pm$0.23} & 4.62{\scriptsize$\pm$0.25} & 1.081 \\
Tango~\cite{ghosal2023tango}               & 866  & 1.62 & 18.1 & 9.4  & 1.40 & 0.667 & 0.121 & 6.00 & 3.69 & 5.19 & 3.93{\scriptsize$\pm$0.27} & 4.74{\scriptsize$\pm$0.19} & 1.402 \\
Tango2~\cite{majumder2024tango}            & 866  & 2.47 & 14.2 & 11.4 & 1.17 & 0.716 & 0.136 & 5.91 & 3.66 & 5.17 & 4.14{\scriptsize$\pm$0.25} & 4.84{\scriptsize$\pm$0.10} & 1.403 \\
TangoFlux~\cite{hung2024tangoflux}         & 515  & 2.20 & 17.3 & 15.0 & 1.20 & 0.728 & 0.142 & 5.89 & 3.69 & 5.10 & 4.11{\scriptsize$\pm$0.22} & 4.88{\scriptsize$\pm$0.11} & 0.123 \\
AudioGen~\cite{kreuk2022audiogen}          & 1500 & 2.01 & 14.4 & 11.6 & 1.54 & 0.631 & 0.118 & 5.28 & 3.15 & 4.55 & 3.99{\scriptsize$\pm$0.28} & 4.57{\scriptsize$\pm$0.26} & 2.010 \\
AudioX~\cite{tian2025audiox}               & 1100 & 2.34 & 14.0 & 13.7 & 1.48 & 0.652 & 0.125 & 5.81 & 3.54 & 5.09 & 3.76{\scriptsize$\pm$0.29} & 4.47{\scriptsize$\pm$0.24} & 0.929 \\
EzAudio~\cite{hai2024ezaudio}              & 874  & 2.55 & 14.2 & 13.0 & 1.22 & 0.692 & 0.139 & 5.55 & 3.45 & 4.82 & 4.22{\scriptsize$\pm$0.19} & 4.77{\scriptsize$\pm$0.18} & 1.209 \\
MMAudio~\cite{cheng2025mmaudio}            & 157  & 3.18 & 12.5 & 13.2 & 1.43 & 0.651 & 0.118 & 5.42 & 3.29 & 4.84 & 4.08{\scriptsize$\pm$0.23} & 4.71{\scriptsize$\pm$0.18} & 0.506 \\
Make-An-Audio 2 \cite{huang2023make2} & 937  & 4.14 & 21.5 & 10.9 & 1.66 & 0.514 & 0.122 & 6.03 & 3.56 & 5.29 & 4.11{\scriptsize$\pm$0.21} & 4.58{\scriptsize$\pm$0.28} & 0.121 \\
StableAudioOpen~\cite{evans2025stable}     & 1210 & 3.70 & 34.9 & 11.8 & 2.26 & 0.404 & 0.100 & 6.61 & 3.28 & 6.24 & 3.85{\scriptsize$\pm$0.26} & 3.86{\scriptsize$\pm$0.39} & 2.701 \\
StableAudioOpen 3~\cite{evans2026stable}       & 2310 & 4.56 & 25.2 & 12.0 & 2.22 & 0.420 & 0.097 & 6.40 & 3.67 & 5.85 & 3.82{\scriptsize$\pm$0.29} & 4.10{\scriptsize$\pm$0.37} & 0.075 \\
AudioLDM2~\cite{liu2024audioldm}           & 1843 & 5.04 & 27.8 & 10.9 & 1.57 & 0.574 & 0.089 & 5.75 & 3.51 & 5.27 & 3.93{\scriptsize$\pm$0.26} & 4.58{\scriptsize$\pm$0.24} & 3.202 \\
Dasheng-AudioGen~\cite{mei2026dasheng}     & 2000 & 3.11 & 22.5 & 9.7  & 1.79 & 0.542 & 0.113 & 5.95 & 3.50 & 5.19 & 4.14{\scriptsize$\pm$0.20} & 4.52{\scriptsize$\pm$0.23} & 0.330 \\
\midrule
\textit{RL Ablations} \\
w/o Semantic Reward   & 234 & 28.83 & 63.8 & 4.4 & 2.91 & 0.298 & 0.065 & 7.28 & 3.74 & 6.70 & 3.78{\scriptsize$\pm$0.24} & 3.80{\scriptsize$\pm$0.43} & 0.32 \\
w/o Aesthetic Reward  & 234 & 2.86 & 19.9 & 9.8 & 1.48 & 0.658 & 0.126 & 5.70 & 3.55 & 5.01 & 4.03{\scriptsize$\pm$0.26} & 4.59{\scriptsize$\pm$0.21} & 0.32 \\
w/o Temporal Reward   & 234 & 2.24 & 19.0 & 9.8 & 1.52 & 0.655 & 0.127 & 5.72 & 3.53 & 5.01 & 4.10{\scriptsize$\pm$0.26} & 4.66{\scriptsize$\pm$0.21} & 0.32 \\
w/o All Rewards       & 234 & 2.61 & 21.9 & 10.0 & 1.59 & 0.635 & 0.125 & 5.61 & 3.43 & 4.94 & 4.09{\scriptsize$\pm$0.24} & 4.58{\scriptsize$\pm$0.25} & 0.32 \\
\midrule
\textbf{VoxAudio (ours)}                   & 234  & 3.47 & 20.5 & 9.5 & 1.47 & 0.657 & 0.126 & 5.78 & 3.59 & 5.12 & 4.09{\scriptsize$\pm$0.24} & 4.79{\scriptsize$\pm$0.17} & 0.32 \\
\bottomrule
\end{tabular}}
\end{small}
\end{center}
\end{table*}

\section{Experiments}
\label{sec:experiments}

\subsection{Implementation Details}
VoxAudio utilizes a pre-trained VAE to encode 24\,kHz audio into 64-dimensional latents at 12.5 frames per second; the backbone comprises 6 joint and 12 fused DiT blocks with hidden size 512, with the full hyperparameter configuration listed in the supplementary material. Training follows two stages. \emph{Supervised flow-matching training}: the model is first trained from scratch on the timed simulation corpus with random chunk boundaries ($p_b{=}0.15$) and EMA (decay 0.9999) for $\sim$195k steps (batch 384, lr $10^{-4}$, cosine decay), and then continues on the blended corpus (70\% simulation, 30\% real narrative) for 60k steps at lr $5{\times}10^{-5}$.  The text condition is dropped with 20\% probability during training. \emph{Preference alignment}: multi-reward NFT at lr $10^{-5}$ with $K{=}6$ rollouts per prompt, reward weights $(w_{\text{CLAP}}, w_{\text{Aes}}, w_{\text{WER}}, w_{\text{TG}}) = (1.0, 0.005, 0.06, 0.05)$, $\beta{=}0.5$, and KL coefficient $10^{-4}$. At inference we use 25 solver steps, text-CFG scale 5.0, chunk size 16, step lag $\Delta{=}5$, and EMA weights.

\subsection{Evaluation Protocol}
We assess audio generation quality with a suite of complementary metrics. \emph{Distributional metrics} include Fr\'echet Distance on VGGish features (FD-VGG) and PANNs features (FD-PANN), Inception Score (ISC), and KL divergence. \emph{Semantic metrics} are measured by PE\textsubscript{AV}~\cite{vyas2025pushingfrontieraudiovisualperception} and LAION-CLAP audio-text similarity, where the CLAP checkpoint used for evaluation is not consistent with the checkpoint employed during reinforcement-learning training. \emph{Speech metrics} are quantified by word error rate (WER) computed with Whisper-large-v3 on the vocal stem isolated by a speech enhancement model. \emph{Aesthetics} are evaluated using Audiobox-Aesthetics~\cite{tjandra2025aes}, including production quality (PQ), content enjoyment (CE), and usefulness (CU). \emph{Temporal grounding} ($\text{TG-IoU}$) measures the interval IoU, $\vert{}I_{\text{det}} \cap I_{\text{inst}}\vert{} / \vert{}I_{\text{det}} \cup I_{\text{inst}}\vert{}$, between the instructed interval $I_{\text{inst}}$ and the speech span $I_{\text{det}}$ detected by PE-A-Frame~\cite{bolya2025PerceptionEncoder} (threshold $0.3$). We further conduct Mean Opinion Score (MOS) tests with 5 listeners. Evaluators rate 50 randomly sampled audio clips per model on a 1–5 Likert scale for overall audio quality (MOS-O) and audio-caption alignment/consistency (MOS-C). Finally, \emph{efficiency} is reported by the real-time factor (RTF) and latency measured on a single A100 GPU.

\begin{figure*}[t]
  \centering
  \includegraphics[width=\textwidth]{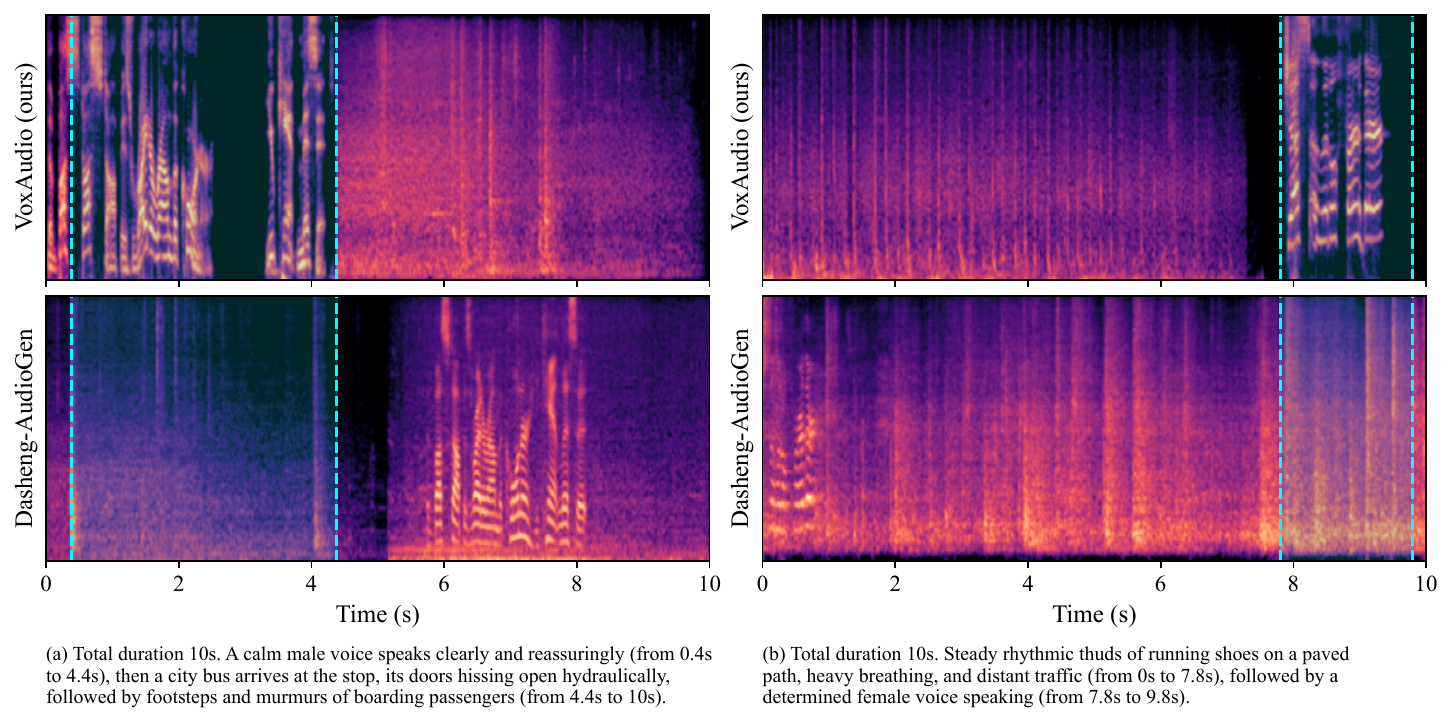}
  \caption{The Generated Mel-spectrograms of VoxAudio (top) and Dasheng-AudioGen (bottom). 
  }
  \label{fig:mel_compare}
\end{figure*}

\begin{table}[t]
\caption{Speech generation on seed-tts-eval (EN).}
\label{tab:seedtts}
\begin{center}
\begin{small}
\setlength{\tabcolsep}{5pt}
\begin{tabular}{lcccc}
\toprule
& & Linguistic & \multicolumn{2}{c}{Efficiency} \\
\cmidrule(lr){3-3} \cmidrule(lr){4-5}
Model & Params(M) & WER$\downarrow$ & Lat.(s)$\downarrow$ & RTF$\downarrow$ \\
\midrule
Qwen3-TTS~\cite{hu2026qwen3}           & 1929 & 1.01\% & 4.82 & 1.54 \\
F5-TTS~\cite{chen2025f5}               & 337  & 1.20\% & 0.70 & 0.28 \\
CosyVoice2~\cite{du2024cosyvoice}      & 639  & 2.16\% & 2.87 & 0.56 \\
CosyVoice3~\cite{du2025cosyvoice}      & 859  & 1.81\% & 2.39 & 0.59 \\
\midrule
Dasheng-AudioGen~\cite{mei2026dasheng} & 2000 & 27.47\% & 3.30 & 0.33 \\
\midrule
\textit{RL Ablations} \\
w/o Linguistic Reward & 234 & 1.78\% & 3.20 & 0.32 \\
\midrule
\textbf{VoxAudio (ours)}               & 234  & 1.61\% & 3.20 & 0.32 \\
\bottomrule
\end{tabular}
\end{small}
\end{center}
\end{table}

\begin{table}[!t]
\caption{Ablation results on VoxBench.}
\label{tab:ablation}
\begin{center}
\footnotesize
\setlength{\tabcolsep}{3pt}
\resizebox{\columnwidth}{!}{%
\begin{tabular}{lcccccc}
\toprule
& Semantic & Linguistic & \multicolumn{3}{c}{Aesthetics} & Temporal \\
\cmidrule(lr){2-2} \cmidrule(lr){3-3} \cmidrule(lr){4-6} \cmidrule(lr){7-7}
Variant & PE\textsubscript{AV}$\uparrow$ & WER$\downarrow$ & PQ$\uparrow$ & CE$\uparrow$ & CU$\uparrow$ & TG-IoU$\uparrow$ \\
\midrule
Default ($C{=}16$, $\Delta{=}5$) & 0.107 & 0.076 & 5.55 & 4.03 & 5.06 & 0.670 \\
Single-chunk & 0.102 & 0.114 & 5.49 & 4.05 & 5.01 & 0.710 \\
\midrule
\multicolumn{7}{l}{\textit{(a) Inference chunk size (Stage 1)}} \\
\midrule
$C=8$  & 0.107 & 0.067 & 5.52 & 4.04 & 5.04 & 0.653 \\
$C=20$ & 0.108 & 0.073 & 5.57 & 4.05 & 5.09 & 0.669 \\
\midrule
\multicolumn{7}{l}{\textit{(b) Streaming step lag (Stage 1)}} \\
\midrule
$\Delta=0$  & 0.105 & 0.177 & 5.62 & 4.06 & 5.07 & 0.682 \\
$\Delta=1$  & 0.107 & 0.113 & 5.61 & 4.07 & 5.09 & 0.680 \\
$\Delta=2$  & 0.108 & 0.084 & 5.60 & 4.08 & 5.10 & 0.676 \\
$\Delta=10$ & 0.107 & 0.070 & 5.54 & 4.06 & 5.08 & 0.656 \\
$\Delta=25$ & 0.107 & 0.121 & 5.45 & 4.00 & 5.00 & 0.670 \\
\midrule
\multicolumn{7}{l}{\textit{(c) Target duration (Stage 2)}} \\
\midrule
10\,s & 0.108 & 0.038 & 5.94 & 4.34 & 5.57 & 0.714 \\
20\,s & 0.080 & 0.127 & 5.59 & 3.88 & 5.19 & 0.536 \\
30\,s & 0.081 & 0.277 & 5.59 & 3.72 & 5.24 & 0.325 \\
\bottomrule
\end{tabular}}
\end{center}
\end{table}

\subsection{Main Results}
\textbf{Unified Vocalized Audio (Table~\ref{tab:unified}).}
To ensure a fair comparison, we evaluate on the English subsets of VoxBench-10s and MECAT. This is because Dasheng-AudioGen generates fixed 10-second clips. We use an LLM to convert our prompts into Dasheng-AudioGen’s standard input format. The results show that our model achieves substantially lower WER: 0.038 and 0.085, compared to 0.264 and 0.151 for Dasheng-AudioGen. We also observe strong performance in temporal controllability. Moreover, although our training does not use MECAT data, our method remains competitive on semantic and aesthetic metrics. On the main human-evaluated MOS metrics, our MOS-O (overall audio quality) is comparable to the baseline, while our MOS-C (audio-caption consistency) is significantly higher. Figure~\ref{fig:mel_compare} visualizes this contrast on an early-instructed and a late-instructed prompt: VoxAudio places articulate speech precisely inside the instructed interval and keeps the ambient bed continuous around it, whereas Dasheng-AudioGen shifts its speech to the wrong part of the clip. 

\textbf{General Audio Synthesis (Table~\ref{tab:audiocaps}).}
As shown in the table, our method demonstrates overall competitive performance against baselines, while achieving marked improvements in WER and temporal alignment. Compared with other purely audio generation models, our streaming generation approach attains competitive results on both main objective and human-evaluated metrics even with a smaller model size and a generation efficiency of around 0.32 RTF.

\textbf{Speech Generation (Table~\ref{tab:seedtts}).}
As shown in the table, on the SEED-TTS-EVAL (EN) benchmark for pure speech generation, our method achieves a WER of 1.61\%, which is substantially lower than 27.47\% for Dasheng-AudioGen, a competing unified model. Moreover, our results are competitive with those of traditional TTS systems.

\subsection{Ablation Study}
\label{sec:ablation}

\textbf{Reinforcement Learning(Table~\ref{tab:ablation}).}
As shown in Tables~\ref{tab:audiocaps}, \ref{tab:seedtts}, and \ref{tab:unified}, we observe that removing a specific reward term in isolation leads to a corresponding degradation in the associated metrics. Moreover, the comparison indicates that the largest impact is on the WER metric. Notably, across different test sets, the results are consistently improved when all reward terms are included.
Additionally, the semantic reward acts as an essential anchor to balance RL training. Without this constraint, optimizing linguistic rewards alone easily leads the model to sacrifice soundscape fidelity and natural intonation, making the RL process highly prone to collapse.

\textbf{Comparison of Different Inference Variants.
(Table~\ref{tab:ablation}).}
To evaluate the model performance, we compare the Stage-1 models under different configurations. From the perspective of different training schemes, we find that using only the non-autoregressive generation with all frames kept under the same noise level achieves better results in the temporal consistency metric, while performing worse than the Flow-AR baseline in other metrics.
Furthermore, under different settings, we show that in inference, using an excessively large or small inference chunk\_size can both lead to worse metrics. Similarly, for the streaming setting, we observe that choosing an overly large or overly small chunk for streaming causes metric degradation, as reflected in the streaming step lag results.

\textbf{Variable-Duration Generation (Table~\ref{tab:ablation}).}
After RL, we test the model under various configurations with target durations of 10/20/30,s (i.e., targets shorter than 10,s, 20,s, and 30,s, respectively). As the target length increases, the overall metrics generally decline. However, for 30,s audio, our WER still reaches the same order of magnitude as the 10,s performance reported by Dasheng-AudioGe. This demonstrates that our model can generate audio of varying lengths effectively.
\section{Limitations and Future Work}
\label{sec:limitations}
While VoxAudio demonstrates strong performance, several key limitations remain. First, the model currently exhibits a mono-lingual bias toward English data and experiences reduced acoustic coherence for audio generation exceeding 30 seconds. Future work will expand VoxCorpus to encompass a broader range of languages and investigate hierarchical architectures to maintain long-range audio consistency. Second, fine-grained control over speaker timbre and emotional prosody is currently limited. To address this, we plan to enhance our three-level paradigm by incorporating timbre and prosody annotations, attribute conditioning channels, and speaker/emotion-alignment rewards into the multi-reward NFT pipeline. Finally, we aim to extend VoxAudio to visual modalities, enabling spatially synchronized audio-visual generation.

\section{Conclusion}

This paper presents VoxAudio, a causal flow-matching framework for controllable, real-time vocalized audio synthesis. By unifying a chunk-agnostic autoregressive architecture with multi-reward Negative-aware Fine-Tuning (NFT), VoxAudio enables low-latency streaming and robust preference alignment across speech and soundscape dimensions. Extensive evaluations on the new VoxCorpus and VoxBench demonstrate its superior performance in speech placement, acoustic quality, and temporal grounding.

\bibliographystyle{IEEEtran}
\bibliography{main}

\appendices
\section{Details of VoxCorpus and VoxBench}
\label{app:data}

\begin{figure*}[htbp]
  \centering
  \includegraphics[width=0.95\textwidth]{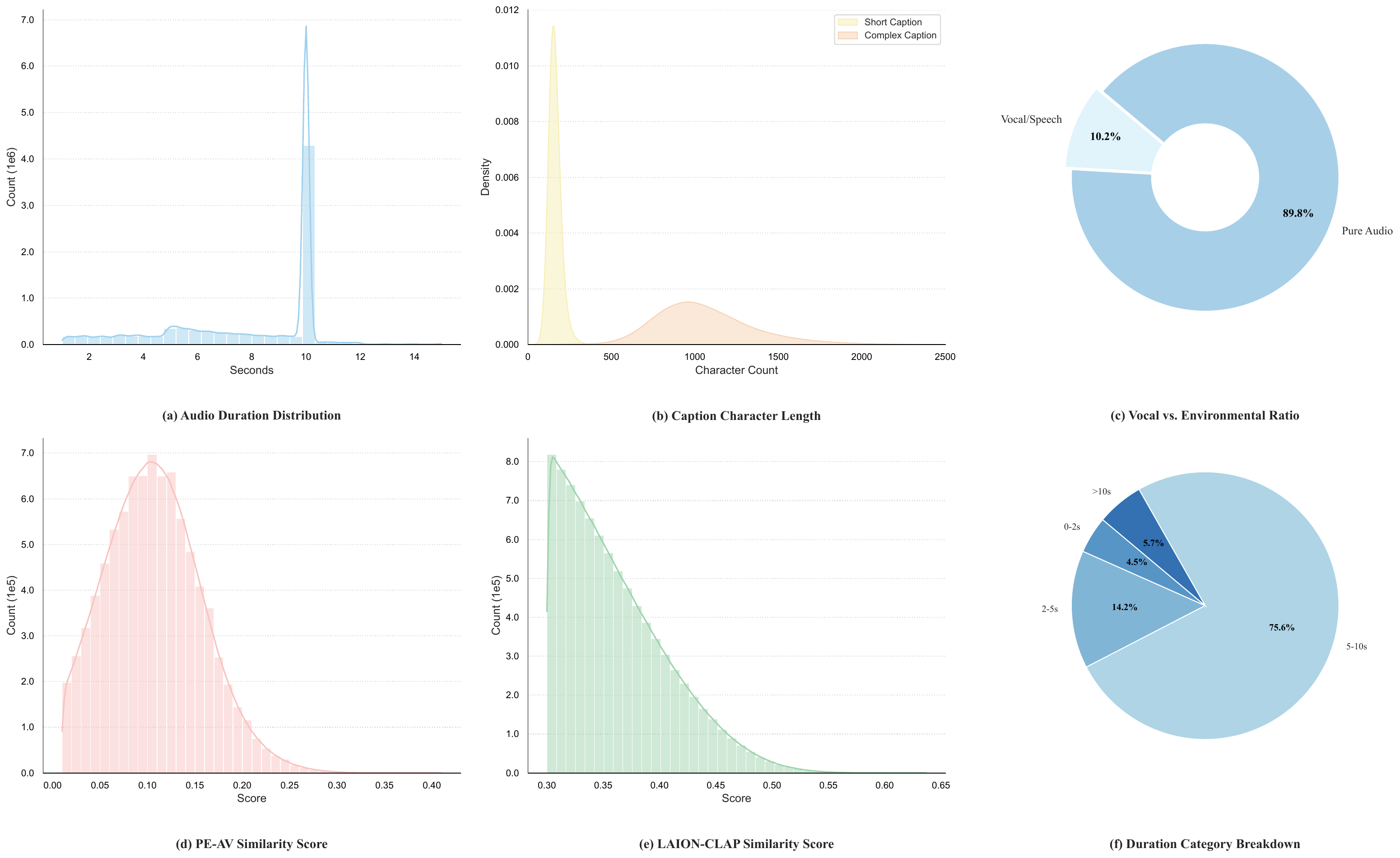}
  \caption{
    Comprehensive statistical analysis of VoxCorpus. 
    (a) Histogram of audio duration. 
    (b) Distribution of caption character lengths for Short and Complex versions. 
    (c) Ratio of audio samples containing speech content. 
    (d) Distribution of PE\textsubscript{AV} similarity scores. 
    (e) Distribution of LAION-CLAP similarity scores. 
    (f) Breakdown of audio durations by category.
  }
  \label{fig:stats_summary}
\end{figure*}

\begin{figure*}[t]
\centering
\begin{promptbox}{Prompt for generating structured audio descriptions}
\small
\textbf{Role:} You are a professional audio description and refinement assistant. Your task is to generate high-quality, detailed English audio descriptions in two versions: \textbf{Short} and \textbf{Complex} for ``Text-to-General-Audio'' tasks.

\textbf{Task Requirements:}
Analyze the provided \texttt{caption} and \texttt{transcription} to generate a JSON object containing:
1. \textbf{Short Version}: Core information refined within strict character limits.
2. \textbf{Complex Version}: Maximum narrative detail without character constraints.

\textbf{Core Rules:} \\
\textbf{1. Strict Adherence:} Descriptions must be extracted directly from the \texttt{caption}. Do not imagine or add details not mentioned in the source. \\[4pt]
\textbf{2. Remove Technical/Electronic Noise:} Prohibit any mention of recording quality, electrical noise, or studio environments (e.g., high-fidelity, reverb, hiss, sample rate). \\[4pt]
\textbf{3. Narrative Sound Effects:} Retain and describe all meaningful sound effects, ambient sounds, and musical instruments (e.g., violin, piano). Preserve timing and dynamic changes. \\[4pt]
\textbf{4. Speech Logic:} 
\begin{itemize}
    \item If Transcription is \texttt{[NONSPEECH]}, remove all vocal descriptions.
    \item If speech exists, replace described content with the exact transcription using the format: \textbf{``[SPEECH\_CONTENT](Transcription)''}. Maintain tone and emotions.
\end{itemize}
\textbf{5. Output Constraints:} \texttt{caption\_short} must be under 200 characters. All placeholder structures must be enclosed in double quotes.

\textbf{JSON Schema:} \\
\{ \\
\hspace*{1em} ``caption\_short'': ``Concise English description (max 200 chars)'', \\
\hspace*{1em} ``caption\_complex'': ``Complete and detailed English description'' \\
\}
\end{promptbox}
\vspace{-2pt}
\caption{Prompt template for generating Short and Complex structured audio descriptions based on raw captions and transcriptions.}
\vspace{-5pt}
\label{app_fig:audio_description_generation}
\end{figure*}

\begin{figure*}[t]
\centering
\begin{promptbox}{Prompt for generating diverse audio scene candidates}
\small
\textbf{Role:} You are a world-class film audio director and sound designer. Based on the provided keywords, conceive \textbf{5 distinct} high-fidelity audio scene candidates.

\textbf{Task Requirements:} \\
\textbf{1. Diversity:} Design 5 different scenarios or perspectives based on the same keyword. \\[4pt]
\textbf{2. Mandatory Vocals:} Each candidate MUST include clear human dialogue or monologue logical to the scene. \\[4pt]
\textbf{3. Caption Consistency:} The \texttt{overall\_caption} MUST explicitly mention the specific lines spoken by characters (e.g., ...someone says: ``Quote''). \\[4pt]
\textbf{4. Duration:} Each scene's total duration must be between 0-10 seconds. \\[4pt]
\textbf{5. Detailed Events:} For vocal events, \texttt{voice\_details} MUST include gender, tone, and emotion. \\[4pt]
\textbf{6. JSON Format:} Return ONLY a JSON object with a root key ``candidates'' containing the list.

\textbf{JSON Schema \& Example:} \\
\{ \\
\hspace*{1em} ``candidates'': [ \\
\hspace*{2em} \{ \\
\hspace*{3em} ``id'': 1, \\
\hspace*{3em} ``overall\_caption'': ``In a forest at dawn... a woman shouts: `Look over there!', followed by rustling leaves.'', \\
\hspace*{3em} ``audio\_event\_timeline'': [ \\
\hspace*{4em} \{ ``start\_time'': ``3.5'', ``end\_time'': ``5.5'', ``event'': ``Speech'', \\
\hspace*{5em} ``voice\_details'': \{ ``gender'': ``Female'', ``emotion'': ``Excited'', ``content'': ``Look over there!'' \} \} \\
\hspace*{3em} ] \\
\hspace*{2em} \} \\
\hspace*{1em} ] \\
\}
\end{promptbox}
\vspace{-2pt}
\caption{Prompt template used for generating diverse, time-aligned audio scene candidates with mandatory vocal content from keywords.}
\vspace{-5pt}
\label{app_fig:keyword_to_caption_generation}
\end{figure*}

\subsection{Real Narrative Corpus: Curation Workflow}
VoxCorpus comprises a timed simulation corpus and a real narrative corpus (Section~\ref{sec:dataset}). This subsection details the curation of the real narrative corpus, whose processing pipeline consists of four primary stages:

\begin{itemize}
    \item \textbf{(1) Sourcing}: We collect high-bitrate audio from extensive web-crawled sources and established open-access datasets, such as AudioSet~\cite{gemmeke2017audio}, VGGSound~\cite{chen2020vggsound}, and LAION-Audio-630K~\cite{laionclap2023}, to ensure comprehensive acoustic diversity across various environments.
    \item \textbf{(2) Segmentation \& Cleaning}: Raw audio is partitioned into segments (shorter than 15\,s) using energy-based audio activity detection. Samples with severe clipping or an excessive noise floor are discarded.
    \item \textbf{(3) Audio Captioning}: We utilize Qwen3-ASR for precise speech transcription and Qwen3-Omni for initial environmental event tagging.
    \item \textbf{(4) Refinement}: A strong LLM fuses these multi-modal labels into coherent ``structured audio descriptions''. 
\end{itemize}

The corpus and benchmark will be released under the \textbf{CC BY-NC-SA 4.0} license for non-commercial research use.

\subsection{Timed Simulation Corpus: Configuration}
The simulation corpus composes clean speech utterances with non-speech recordings under five layout modes (speech-only, sound-only, speech-then-sound, sound-then-speech, and overlay). The mixer randomizes the speech onset within the clip, the speech-to-background signal-to-noise ratio, and an optional trailing-silence segment. Captions are rendered from the composition parameters: transcripts are quoted verbatim, segment intervals are annotated, and the total-duration prefix, segment times, and numeric precision are stochastically dropped or varied so that the model accepts captions at any annotation density.

\subsection{Dataset Statistics}
We perform a rigorous statistical analysis of the curated corpus, which yields a total of 879,768 samples after filtering. The overall distributions are visualized in Figure~\ref{fig:stats_summary}.

\textbf{Duration Characteristics}: As illustrated in Figures~\ref{fig:stats_summary}(a) and (f), the audio samples are primarily distributed between 5s and 10s, which is well-suited for modeling complex soundscapes.

\textbf{Caption Diversity}: Figure~\ref{fig:stats_summary}(b) highlights the contrast between our dual-caption system.  Captions provide concise summaries of sound sources and locations, with lengths typically capped at 500 characters.

\textbf{Speech Content}: According to Figure~\ref{fig:stats_summary}(c), 10.2\% of the dataset contains intelligible human speech integrated with environmental sounds. This specific subset enables the model to learn the intricate balance between vocal clarity and background ambiance.

\textbf{Semantic Alignment Quality}: Figures~\ref{fig:stats_summary}(d) and (e) present the similarity score distributions for PE\textsubscript{AV} and LAION-CLAP, respectively. Notably, the distributions are truncated at the lower end; this is a consequence of our strict data pruning strategy, where samples falling below a predefined similarity threshold are excluded to maintain high-fidelity text-audio alignment.

\subsection{VoxBench Details}
\label{app:bench}
VoxBench is constructed by sampling from a library of fine-grained keywords across 12 acoustic domains (e.g., Emergency, Cafe, Office, Storm) and prompting an LLM with the template in Figure~\ref{app_fig:keyword_to_caption_generation} to generate time-aligned captions with mandatory speech components. After consistency filtering and manual review, the evaluation set contains 977 held-out items. The reinforcement-learning prompt set contains $\sim$4.3k items with variable target durations (10--30\,s) and is deduplicated against the evaluation set at both the quote and the scene level. The variable-duration suite contains 450 additional items in three tiers (10/20/30\,s).
\section{Model Architecture Details}
\label{app:model}

We summarize the key hyperparameters of the VoxAudio architecture in Table~\ref{tab:model_params}. The following subsections provide a detailed breakdown of our model's core components and the implementation of our streaming mechanism.

\begin{table}[h]
\caption{Hyperparameters of the VoxAudio backbone.}
\label{tab:model_params}
\begin{center}
\begin{small}
\begin{tabular}{lc}
\toprule
Parameter & Value \\
\midrule
Joint Blocks & 6 \\
Fused Blocks & 12 \\
Hidden Dimension $D$ & 512 \\
Attention Heads & 16 \\
MLP Ratio & 4.0 \\
Chunk Size $C$ (default; swept 8--20) & 16 \\
Kernel Size & 7 \\
Text Encoder Dimension & 1024 \\
Latent Dimension & 64 \\
\midrule
Total Trainable Parameters & 234M \\
\bottomrule
\end{tabular}
\end{small}
\end{center}
\end{table}

\subsection{Latent Space Representation (VAE)}
VoxAudio operates in a compressed latent space provided by the pre-trained Universe audio VAE, which is trained on a broad mixture of speech, sound effects, and music. The VAE maps a 24kHz waveform $x$ into a 64-dimensional latent $\mathbf{z}$ at 12.5 frames per second and is kept frozen throughout all training stages. Its reconstruction objective combines a multi-resolution STFT loss with a GAN-based adversarial loss to ensure high-fidelity signal recovery.

\section{Experimental Details}
\label{app:exp}
\subsection{Training Implementation}
\label{app:training}

Training follows the two-stage pipeline of Section~\ref{sec:training}. The \emph{supervised flow-matching stage} first trains the model from scratch on the timed simulation corpus with random chunk boundaries ($p_b{=}0.15$), EMA (decay 0.9999), and duration conditioning, for $\sim$195k steps at batch size 384 (lr $10^{-4}$, cosine decay), and then continues training on the blended corpus (70\% simulation, 30\% real narrative) for 60k steps at lr $5\times 10^{-5}$; the streaming chunk size is then a free parameter at the inference stage, and for the training-granularity ablation we additionally train a single-chunk (non-autoregressive) variant. The \emph{preference-alignment stage} performs multi-reward NFT for 300 optimization steps at lr $10^{-5}$ with $K{=}6$ CFG-guided rollouts per prompt (288 rollouts per update at guidance scale 5.0) over variable-duration prompts (10--30\,s), using $\beta{=}0.5$, a KL coefficient of $10^{-4}$ toward a frozen reference, and a softly-updated rollout anchor. Supervised runs use 8 NVIDIA A100 GPUs per variant; the RL stage uses 8 GPUs.

\subsection{Evaluation Metrics Formulation}
\textbf{Semantic Similarity.} We employ Cosine Similarity to compute CLAP and PE\textsubscript{AV} scores. Given audio embedding $e_a$ and text embedding $e_r$, the score is:
\begin{equation}
S_{\text{cos}}(e_a, e_r) = \frac{e_a \cdot e_r}{\|e_a\| \|e_r\|}.
\end{equation}

\textbf{Temporal Grounding (TG-IoU).}
We localize speech in the generated audio with the frame-level audio-text grounding model PE-A-Frame~\cite{bolya2025PerceptionEncoder}, using the query ``a person speaking'' with a detection threshold of 0.3. Frames whose scores exceed the threshold are merged into a detected span $I_{\text{det}}$, and TG-IoU is computed against the instructed interval $I_{\text{inst}}$ as:
\begin{equation}
\text{TG-IoU} = \frac{|I_{\text{det}} \cap I_{\text{inst}}|}{|I_{\text{det}} \cup I_{\text{inst}}|}.
\end{equation}
When a caption specifies multiple speech intervals, each instructed interval is matched to the detected span with the highest overlap and the scores are averaged.

\textbf{Word Error Rate (WER).}
The vocal stem is transcribed by Whisper-large-v3; after text normalization, the transcription is scored against the transcript that each system consumed in its prompt.

\textbf{Inference efficiency.}
Inference efficiency, including latency and the Real-Time Factor (RTF), is measured with the official implementation of each system on a single NVIDIA A100 GPU. All measurements use a batch size of 1, and the reported values are averaged over 100 independent generation trials.

\textbf{Mean Opinion Scores.}
We evaluate subjective performance using Mean Opinion Scores (MOS). For each test, 50 audio clips per model are randomly sampled from the evaluation set. A pool of 16 independent evaluators participated in the assessment, ensuring each audio clip received at least five ratings on a 5-point Likert scale (1 = Poor, 5 = Excellent) across two dimensions: MOS-O (Overall Audio Quality, evaluating perceptual clarity, absence of artifacts, and acoustic naturalness) and MOS-C (Audio-Caption Consistency, assessing text-audio alignment across speech content, temporal timing, and background soundscape). All participants completed the evaluation using standard headphones in a quiet environment.

\subsection{Baseline Implementation}
All baselines are executed with their official inference configurations. For diffusion-based text-to-audio systems, we employ the respective official solvers, generation steps, and classifier-free guidance scales; for the discrete autoregressive baseline, we adopt the sampling configuration of its original implementation. TTS systems receive the official reference utterances of seed-tts-eval under the voice-cloning protocol, whereas VoxAudio synthesizes from text alone without any reference audio. On MECAT-en, every system consumes captions in its native input format.

\subsection{Inference Efficiency Profile}
\label{app:efficiency}
Table~\ref{tab:efficiency} reports the efficiency counterpart of the inference-configuration: total latency, RTF, and first-chunk latency (the wall-clock time until the first audio chunk is emitted), measured on a single A100 (batch size 1, NFE 25, text-CFG 5.0) and averaged over 100 trials. One latent frame spans 80\,ms, so a chunk of size $C$ emits $0.08C$ seconds of audio. Three observations follow. First, RTF decreases roughly inversely with the chunk size, while the first-chunk latency stays constant, so $C$ trades streaming granularity against throughput. Second, the step lag $\Delta$ governs the latency--quality trade-off: $\Delta{=}0$ reaches a 0.15\,s first response but degrades linguistic accuracy, whereas $\Delta{=}25$ approaches fully sequential generation with RTF above 1. Third, the single-chunk variant has the lowest total compute but must synthesize the entire clip before emitting any audio, and the RTF of the default configuration remains stable at longer target durations (0.34 at 20\,s, 0.33 at 30\,s), reflecting the constant per-chunk cost of the sliding-window attention with KV caching.

\begin{table}[t]
\caption{Inference efficiency on a single A100 averaged over 100 trials.}
\label{tab:efficiency}
\begin{center}
\begin{small}
\setlength{\tabcolsep}{5pt}
\begin{tabular}{lccc}
\toprule
Configuration & Latency (s)$\downarrow$ & RTF$\downarrow$ & First-chunk (s)$\downarrow$ \\
\midrule
Default ($C{=}16$, $\Delta{=}5$) & 3.20 & 0.32 & 1.38 \\
Single-chunk & 1.63 & 0.16 & 1.63 \\
\midrule
\multicolumn{4}{l}{\textit{(a) Inference chunk size ($\Delta{=}5$)}} \\
\midrule
$C=8$  & 5.18 & 0.52 & 1.37 \\
$C=20$ & 2.92 & 0.29 & 1.37 \\
\midrule
\multicolumn{4}{l}{\textit{(b) Streaming step lag ($C{=}16$)}} \\
\midrule
$\Delta=0$  & 1.38 & 0.14 & 0.15 \\
$\Delta=1$  & 1.81 & 0.18 & 1.39 \\
$\Delta=2$  & 2.13 & 0.21 & 1.40 \\
$\Delta=10$ & 4.98 & 0.50 & 1.40 \\
$\Delta=25$ & 10.18 & 1.02 & 1.34 \\
\midrule
\multicolumn{4}{l}{\textit{(c) Target duration ($C{=}16$, $\Delta{=}5$)}} \\
\midrule
20\,s & 6.81 & 0.34 & 1.84 \\
30\,s & 9.96 & 0.33 & 1.88 \\
\bottomrule
\end{tabular}
\end{small}
\end{center}
\end{table}

\subsection{MECAT Per-Category Results}
Table~\ref{tab:mecat_cat} reports per-category results on MECAT-en, following the category taxonomy of~\cite{mei2026dasheng}: S/M/A denote the presence of speech, music, and sound effects, with counts taken over the quality-filtered subset. Distributional metrics favor Dasheng-AudioGen (FD-VGG 1.51, FD-PANN 9.88, KL 0.75, ISC 4.74) across categories, which is expected since MECAT is drawn from its own training domain while VoxAudio never sees MECAT-domain audio. Conversely, VoxAudio roughly halves WER in every speech-bearing category (e.g., 0.229$\to$0.104 on speech-only clips), confirming that the linguistic advantage is uniform rather than driven by a single content type.

\begin{table}[t]
\caption{Per-category results on MECAT-en.}
\label{tab:mecat_cat}
\begin{center}
\begin{small}
\setlength{\tabcolsep}{3pt}
\resizebox{\columnwidth}{!}{%
\begin{tabular}{lccccccc}
\toprule
& & \multicolumn{2}{c}{FD-VGG$\downarrow$} & \multicolumn{2}{c}{KL$\downarrow$} & \multicolumn{2}{c}{WER (en)$\downarrow$} \\
\cmidrule(lr){3-4} \cmidrule(lr){5-6} \cmidrule(lr){7-8}
Category & $n$ & Dasheng & \textbf{VoxAudio} & Dasheng & \textbf{VoxAudio} & Dasheng & \textbf{VoxAudio} \\
\midrule
Overall                   & 2621 & 1.51 & 4.16 & 0.75 & 1.33 & 0.151 & \textbf{0.085} \\
\midrule
S00 (speech)              & 393 & 2.11 & 6.88 & 0.42 & 1.03 & 0.229 & \textbf{0.104} \\
SM0 (speech+music)        & 397 & 2.26 & 6.78 & 0.38 & 0.78 & 0.110 & \textbf{0.069} \\
S0A (speech+sfx)          & 390 & 2.06 & 6.63 & 0.65 & 1.43 & 0.117 & \textbf{0.089} \\
SMA (speech+music+sfx)    & 263 & 2.52 & 6.33 & 0.63 & 1.48 & 0.152 & \textbf{0.079} \\
0M0 (music)               & 400 & 2.16 & 3.14 & 0.69 & 0.86 & ---   & --- \\
00A (sfx)                 & 400 & 4.61 & 9.68 & 1.47 & 1.80 & ---   & --- \\
0MA (music+sfx)           & 199 & 3.83 & 5.56 & 1.34 & 1.77 & ---   & --- \\
000 (other)               & 179 & 0.70 & 11.80 & 0.55 & 2.35 & ---   & --- \\
\bottomrule
\end{tabular}}
\end{small}
\end{center}
\end{table}

\end{document}